# Adhesion mechanism of temperature effects on Sn coating on the carbon fiber reinforced polymer substrate by cold spray

Jiayu Sun[a,b], Shaoyun Zhou[c], Kenta Yamanaka[b,*], Hukang bina[b], Yuji Ichikawa[c], Hiroki Saito[c], Kazuhiro Ogawa[c], Akihiko Chiba[b]

[a]Department of Materials Processing, Graduate School of Engineering, Tohoku University, Japan

[b]Institute for Materials Research, Tohoku University, Japan

[c]Fracture and Reliability Research Institute, Tohoku University, Japan

*Corresponding author (k_yamanaka@imr.tohoku.ac.jp)

## Abstract

Metalization of carbon fiber reinforced polymers (CFRPs) composites by the surface modification method to enhance their electrical conductivity, thermal conductivity, electromagnetic shielding, erosion, and radiation protection, has a significant meaning in the aerospace field. In this study, Sn coating was successfully fabricated on the CFRP composite substrate via low-pressure cold spray under four gas temperatures (473K, 523K, 573K, and 623K). Their bonding mechanism was explored via the surface observation after peel-off adhesion strength, accompanying with surface temperature distribution investigation. The results indicates that we cannot obtain coating at 623 K, the epoxy matrix of the CFRP substrate was gradually eroded during deposition over 523 K. Meanwhile, Sn particles melt under 623 K condition. Three kinds of interfaces: Sn/epoxy, Sn/CF, and Sn/CF/epoxy are revealed as characteristics with respect to different gas temperatures to explore the bonding mechanisms.

## 1. Introduction

Carbon fiber reinforced polymers (CFRPs) have been increasingly used in industries, such as the aerospace and automotive sectors, owing to their lightweight potential. However, the low thermal and electrical conductivity and low humidity resistance of the epoxy matrix in CFRP are fatal weaknesses [1]. Notably, in aircraft parts, joining between metal and CFRP surface is crucial for fuselage from a lightning strike, joining between metal alloys and CFRP of internal parts can take into account the lightweight and enhance the surface protection simultaneously [2,3]. Therefore, the metallization of the CFRP surface has attracted more interest [4].

Generally, lots of conventional methods have been developed for metallization, such as thermal spray and high-velocity oxy-fuel (HVOF) spraying process, which are conducted at elevated temperatures (e.g., over 1273K) and cannot avoid thermal effects on the epoxy matrix to guarantee the coating quality [5,6]. Other methods, such as flame spraying, electric arc wire spraying, and plasma spraying, cannot avoid the degradation and decomposition of the polymer matrix caused by their higher energy input [5,7]. Although physical and chemical vapor deposition (PVD and CVD) enable atomic-level controllable deposition procedures and coating thickness, they are limited by the maximum dimension of processable workpieces and costs [8,9]. The epoxy matrix in CFRP has a low thermal resistance and brittle nature; it will thus carbonize and degrade during the process, which leads to erosion and makes the coating difficult. Compared to other techniques mentioned before, cold spray offers some intriguing advantages and promising for the metallization of epoxy-based CFRP [10]. Metal particles sprayed from the nozzle are deposited on the substrate via plastic deformation at a relatively low temperature and pressure; thus, the surface heat damage can be minimized [11–15]. This technique has been employed in the metal coating deposition on CFRPs [1,2,16,17].

Sn is potential as a coated metal material with a long history of use due to its soft nature, low melting temperature (505 K), and successful precedent [2]. Also, because metal Sn has good stretch properties, it is not easy to oxidize in the air; its alloys have anti-corrosion properties, so often be used as an anti-corrosion layer for other metals [18]. Up until recently, only a few researchers have explored the CFRP (epoxy matrix) metallization by cold spray. Che *et al.* fabricated non-continuous tin coatings on the CFRP surface and proposed the "crack filling mechanism" to explain the reason why soft coatings can be deposited on the eroded surface. The effects of particle velocity and gas temperature on the coating quality were analyzed to reveal the processing window [2]. Powder melting phenomenon has been observed during the process verified by the numerical simulation [19]. Ganesan *et al.* investigated coating behavior in the cold spray of tin and copper on CFRP in the various combinations of pressure and temperature and found that Sn and Cu particles do not deform against the soft polymer substrate, leading to poor coating adhesion strength [17]. However, the bonding mechanism of Sn/CFRP, specifically, the Sn/epoxy and Sn/Carbon fiber (CF) joint; bonding energy comparison corresponding to vary temperature values has not been clarified yet.

The aim of this study is to investigate the Sn coating metallization bonding mechanism on the CFRP under various gas temperature conditions during cold spray, exploring and classifying deposition performance, analyzing the relationship of temperature and erosion condition, joint behavior under increasing process temperature conditions.

## 2. Materials and methods

### 2.1. Sample preparation

In this study, water atomized tin powder (Sn-AtW-250, Fukuda Metal Foil & Powder, Japan) was used as a feedstock material with a mean size of 24.28um. The particles were non-spherical, irregular (rod-like) water atomization Sn powder. The particle size distribution was characterized by using Laser light scattering diffraction particle size analyzer (LS230, Beckman Coulter, USA). The morphology and diameter distribution of the tin powder is shown in Fig. 1(a) and 1(b). The CFRP plates, which consisted of epoxy matrix and carbon fiber (CF) reinforcements, were provided by Jamco, Japan. Each laminate was made of 52 plies ($[90/0]_{2s}$); two adjacent layers had perpendicular fiber directions shown in Fig. 1(c). The plate was cut into small pieces with a dimension of $30 \times 30 \times 2$ mm$^3$. The CFRP samples were degreased with acetone without sandblasting pretreatment before subjected to the cold spray experiments.

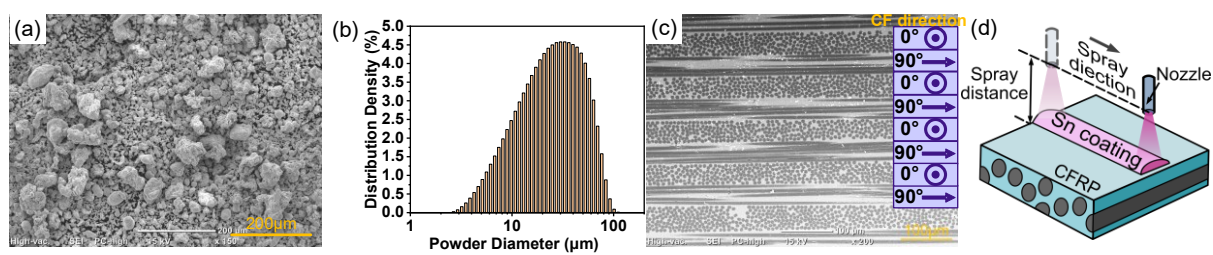

**Figure 1**. Feedstock powder condition (a) SEM image of powder, (b) particle size distribution of the tin powder, (c) CFRP sample cross-section figure, and (d)

schematic of spray distance.

The cold spray experiments were conducted using a commercial low-pressure cold spray system Dymet 423J (Obninsk Centre for Powder Spraying, Russia). Compressed air was heated and delivered through a converging-diverging (de Laval) nozzle. A single-layer deposition along a single spray path was performed to testify the interfacial performance of coating and substrate. Meanwhile, the multi-layer (48 coating layers) experiments were conducted to get a thicker coating to investigate the bonding mechanism by the pull-off test. The processing parameters for the cold spray experiments are listed in Table 1. The schematic figure of spray distance is shown in the Fig. 1(d). Every experiment under different temperature conditions was repeated three times to eliminate system error and measurement deviations.

**Table 1.** Cold spray conditions employed in this study.

| The number of passes | Gas pressure (MPa) | Gas temperature (K) | Spray distance (mm) | Gun traverse speed (mm/s) | Powder feed rate (g/min) |
|---|---|---|---|---|---|
| 1, 48 | 0.4 | 473, 523, 573, 623 | 20 | 30 | 7.0 |

*2.2. Characterization of the cold sprayed materials*

Microstructural characterizations are measured by field-emission scanning electron microscopy (FESEM, SU-70, Hitachi, and S-3400N, Hitachi, Japan) at 15KV. The energy dispersive X-ray analysis (EDX) was conducted on benchtop SEM (JSM-7500F, JEOL, Japan) at 15KV of acceleration voltage. It should be noted that epoxy can be marked by Aluminum element due to elemental introduction by pretreatment during the CFRP manufacturing period, which facilities us to trace epoxy. The measurement of coating thickness, erosion depth, and surface roughness were observed using a digital microscope (VHX-5000. Keyence, Japan). Also, the surface roughness is as consistent as possible to alleviate system error and testing deviation, its mean value of arithmetical mean height (Sa) is 34.02±5.3 μm. Samples for cross-section microstructural analyses were ground with emery paper from 500 to 2000 grade, polished with 1-μm and 0.04-μm suspension subsequently (OP-S, Struers, Japan).

*2.3 Surface temperature measurements*

the InfRec R300 high-resolution infrared thermography camera (NEC Avio Infrared Technologies, Japan) was installed in front of the shifting stage to ensure the same solution under four gas temperatures. The measurement range was set as 273 K to 773 K, ambient temperature, and humidity were set as 298 K and 46%, respectively. InfReC Analyzer NS9500 Professional software was employed to analyze the raw data captured every second. The emissivity value of CFRP and Sn (unoxidized) was set as 0.8 and 0.05 at the temperature ranging from 296 K to 623 K, respectively [20,21].

*2.4 Peel-off adhesion strength test*

The pull-off adhesion strength test was tested three times to eliminate the deviation, conducted by Elcometer 106 pull-off adhesion tester (Elcometer, UK) with a 50mm diameter dolly, following ISO 4624 and ASTM D4541 national and international standards. Before the pull-off test, the samples were prepared by global coverage of Sn coating, the overlap under 473 K and 523 K is 1 mm, and 2 mm for cases under 573 K and 623 K. Then samples were polished to the same thickness roughly, the whole thickness contains the same substrate thickness is around 2.10±0.3 mm.

# 3. Results

*3.1. Surface morphology analysis*

Figure 2 (a) and (b) shows the SEM images of the surface morphology of the single-layer Sn coating obtained at various gas temperature conditions. With increasing the gas temperature, the coating becomes sparser, the epoxy matrix degraded and CFs were exposed when the temperature reaches 573K especially; thus, Sn particles directly deposited on the CFs of the exposed first laminate. It is important to note that the whole laminate erosion consists of epoxy degradation and decomposition, CF breakage and fracture. Moreover at 673 K, two continuous CF laminates were exposed by erosion subsequently due to stronger impact effects, the broken and twisted fibers were observed at the interface of two layers. It is easy to identify because CFs in two continuous laminates are mutually perpendicular. According to previous research, the increased pressure and temperature result in increasing particle velocity, the localized melting zone increases; on the contrary, the stored elastic energy decreases.[22,23] As a consequence, even for soften, Sn particles can penetrate the substrate rather than rebound. [24]

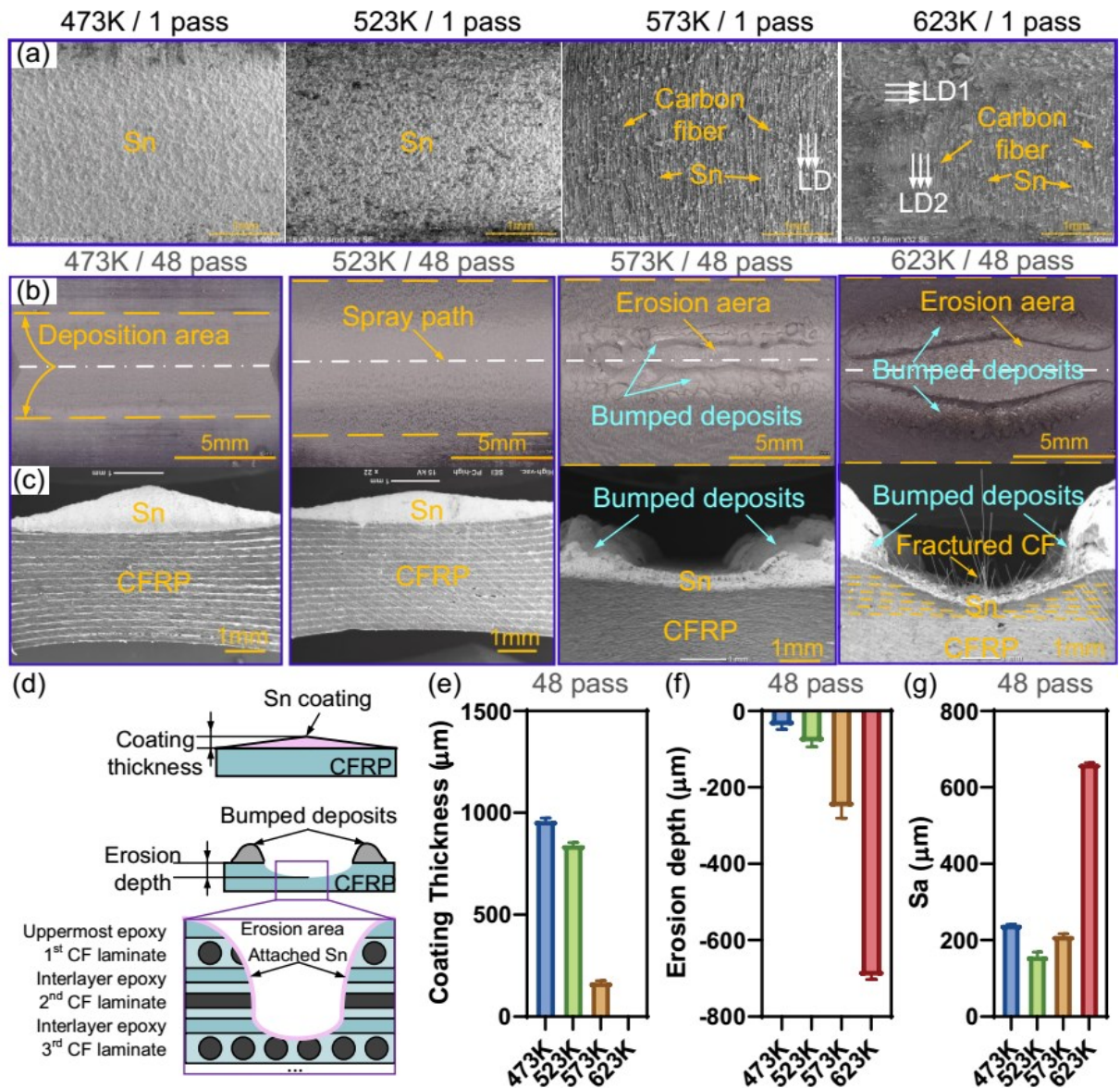

**Figure 2**. Morphology of coatings under four temperature conditions (a) high magnification SEM observation of the 1-pass samples, (b) surface topography at low magnification of the 48-pass conditions, (c) cross-section figure of the 48-pass conditions, (d) Schematic figure of erosion behavior, (e) coating thickness, (f) erosion depth, and (g) surface roughness (Sa).

The enlarged surface topology characterizations can be investigated in Fig. 2(a). The surface of the Sn layer obtained at 473K exhibits thicker and denser compacted coatings. While the gas temperature increases to 523 K, the sparser coating with partial epoxy explosion can be identified instead. Moreover, surface erosion condition became severer when increasing the gas temperature from 523K to 573K and 623K. Consequently, the epoxy matrix gradually decomposed, accompanying with CFs of the first laminate exposure and fracture, CFs get a chance to contact with Sn particles directly under 573K. Under 623K condition, more opportunities occur to make it easier to trigger Sn/CF joint rather than Sn/epoxy joint due to the gradually disappearing uppermost epoxy. After the erosion of the first laminate, the epoxy part between the first and second laminates was eroded immediately, then following with the erosion of the second laminates. It is worth mentioning that erosion was observed within two laminates under all gas temperature conditions.

After 48 passes deposition, the surface and cross-section performances were characterized by an Optical microscope shown in Fig. 2(b), with the temperature increasing from 473 K to 523 K, the coatings are gradually getting thinner and sparser. Specifically, the deposition area (marked within two yellow dotted lines) along the spray path (marked by white central line) becomes larger at the same scale, indicates the coating coverage area increases [25,26]. Though the single-pass width is larger than the nozzle diameter theoretically, increased temperature makes the tendency more dramatic with a larger coating width, leading to coating thickness decrease [27]. Apparently, no coating was obtained along with the spray path under 623 K due to severe erosion. However, on the edge of the erosion-formed valley, some bumped deposits formed, probably caused by beam-edge sputtering accompanying by partial particle melting. Figure 3(d) shows the schematic illustration of the coating thickness and erosion depth defined in this study. The coating thickness and erosion thickness were directly determined from the cross-sectional observations by measuring the height above the initial substrate surface and the depth from the initial substrate surface, respectively. Consequently, as elucidated in Fig. 2(e), 2(f) for the 48-pass condition, with increasing temperature, coating thickness decreases, no coating obtained under 623K; deposition areas increases, erosion depth increases, one laminate was eroded under 573K, two continuous laminates were eroded under 623K. Arithmetical mean height (Sa), which expresses as an absolute value of the height of each measurement point compared to the arithmetical mean height of the surface [28], was used here. The exist of thicker coating at 473 K and bumped deposits at 573 K, and 623 K make the Sa value higher than sparse coating at 523 K, the highest value can be obtained at 623 K.

*3.2. Single-layer interfacial bonding behavior*

In Fig. 3(a), we define defects into three types. The slits inside the epoxy or coating are regarded as crack, defects located in the interface can be noted as gaps, voids can be regarded as a bigger vacancy.

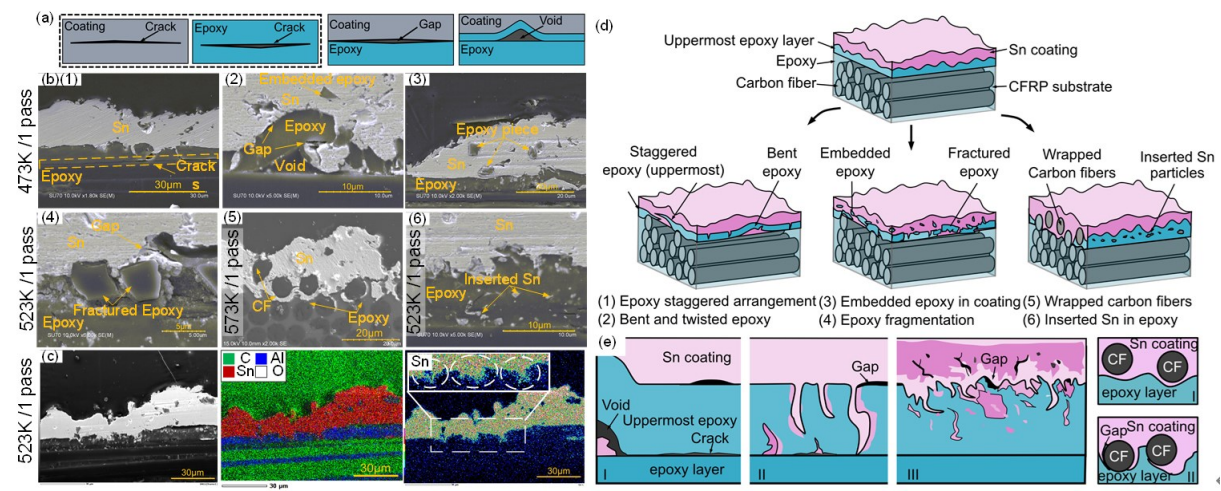

**Figure 3**. Cross-section figure of the interface performance (a) schematic of classification of defects, (b) interfacial behavior of coating under 473K, 523K, and 573 K, (d) elemental distribution at the interface obtained at 523 K by EDX, (e)

schematic figure of six patterns of the interface, and (b) schematic figure of classification of the interface.

The interface behaviors between the Sn coating, the uppermost epoxy, and exposure CFs of the CFRP substrate are illustrated in Fig. 3(b). Their corresponding schematic figures are shown in Fig. 3(d). There exist three kinds of interfaces under 473K and 523 K. Specifically, the first one has a distinct and differentiated dividing interface; the other one is an interlocked interface charactered by the staggering performance at the intertwined dividing interface, the last one describes the epoxy pieces embedded in the Sn coating shown in Fig. 3(b-1, 2, 3). Apparently, their common point is clear Sn/epoxy boundaries without involving of CFs. Moreover, these distinct boundaries at the interface of Sn/epoxy are occasionally accompanied by gaps illustrated. Interestingly, the surficial epoxy part bent but not broke under 473K, epoxy experiences fracture, and a vacancy generated. The deposited Sn partially fulfilled such voids, and some Sn penetrated the epoxy part. Also, apparent cracks inside the coating can be seen, which would weaken the bonding strength.

Figure. 3 (b-4, 6) represents the interface characteristics under 523K, while Fig. 3 (b-5) under 573 K. Notably, broken epoxy pieces result from increased temperature can be observed; they emerged in the moment of consolidation of the coating before degradation, some tiny Sn particles embedded inside the epoxy. Figure 3(c) shows the EDX element distribution map; the uppermost epoxy parts show a staggered arrangement with inserted Sn parts. At 573 K, the thinnest coating can be achieved due to erosion occurrence. Exposed CFs get a chance to trigger Sn/CF joint with the features of CFs wrapped and surrounded by Sn coating.

Overall, the interface performances of Sn coating and CFRP substrate can be briefly classified into six patterns illustrated in Fig. 3(d). Also, three types of Sn/epoxy interfaces, two types of Sn/CF and Sn/CF/epoxy interfaces, can be summarized in Fig. 3(e). As for Sn/epoxy joint, the interface has a clear line of demarcation shown in type I, commonly occurs at 473K. Type II describes the uppermost layer of epoxy is inserted vertically by coating; this performance usually can be seen under 523K. Concerning type III, an interlocked boundary can be observed, and the interface jagged and hooked, the coating and substrate staggered and penetrated into each other, similar to the mechanical interlocking bonding mechanism, anchoring effects of the coating can enhance its adhesion strength [4]. With respect to the Sn/CF, fully surrounded CFs is the representative features of type I. Instead, type II shows Sn/CF/epoxy joint with three interfaces lead to a complicated situation.

*3.3. Multi-layer interfacial bonding behavior*

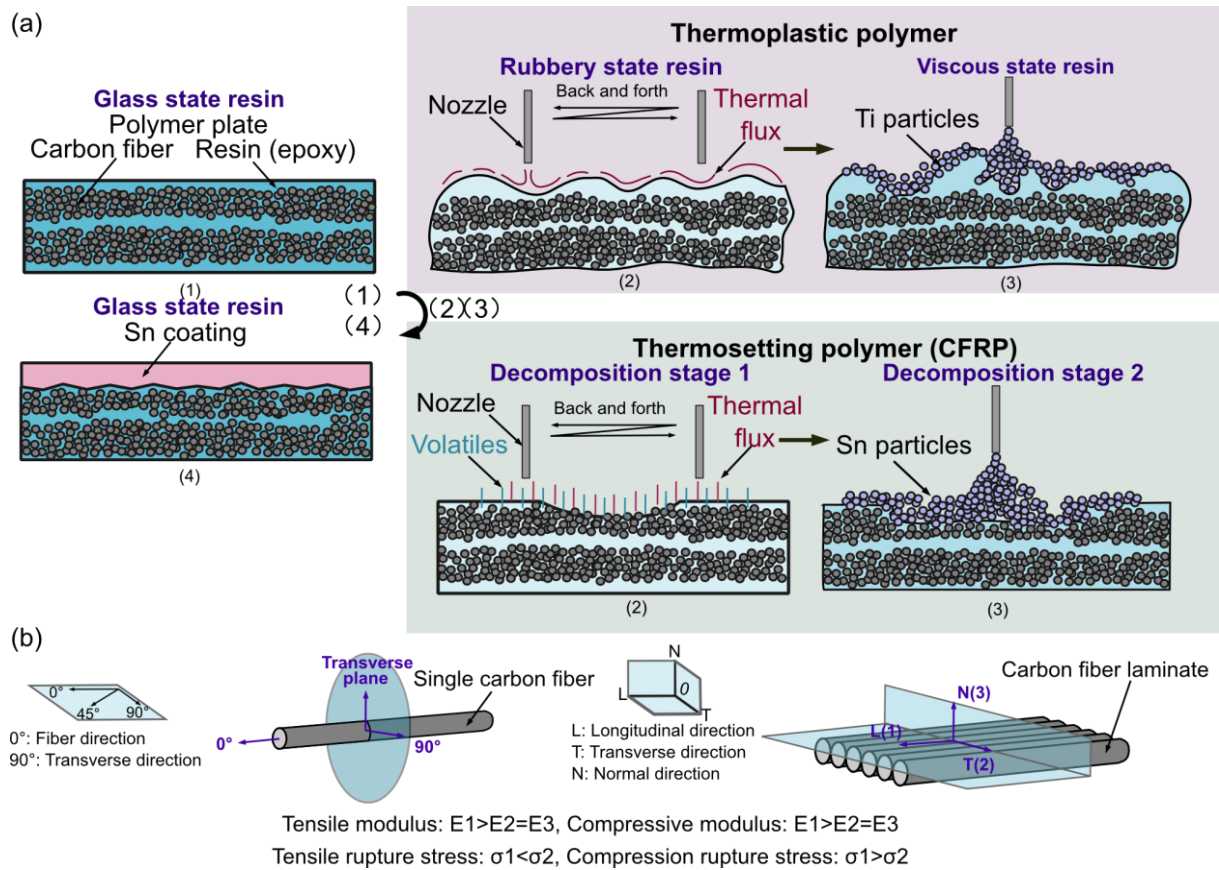

**Figure 4**. Schematic of epoxy and CF behavior during the process (a) transformation procedure of thermosetting polymer under higher temperature, (b) Force model of single carbon fiber and laminate, and (c) carbon fragment performances.

When the CFRP samples under the high-temperature circumstance such as 573K and 623K, as shown in Fig. 4(a), epoxy role as thermosetting polymer would fracture first, then degrade and decompose during the process, if exclude inserted fractured epoxy, they experience incompletely because insertion in a coating prevents further degradation. Not similar to thermoplastic polymers, such as polyetheretherketone (PEEK) or Poly Phenylene Sulfide Resin (PPS), epoxy does not go through a rubber state after heating. In general, after the degradation of epoxy resin on the uppermost surface, the first CF laminate will be exposed, which provides accessible opportunities for the deposition of Sn particles on CF directly. However, the high impact energy would directly transfer from particles to fibers and result in CF fracture. As shown in Fig. 4(b), single carbon fiber and one-layer laminate are anisotropic materials [29]. In the fiber direction, it has much higher tensile modulus and compressive modulus, but tensile rupture stress and compression rupture stress are higher along the transverse direction [30,31]. The impact direction of the powder is perpendicular to the CF, coincides with normal direction, fibers are more easily compressed, then fractured into fragments, leading to deformation, crack generation, and fracture [32,33].

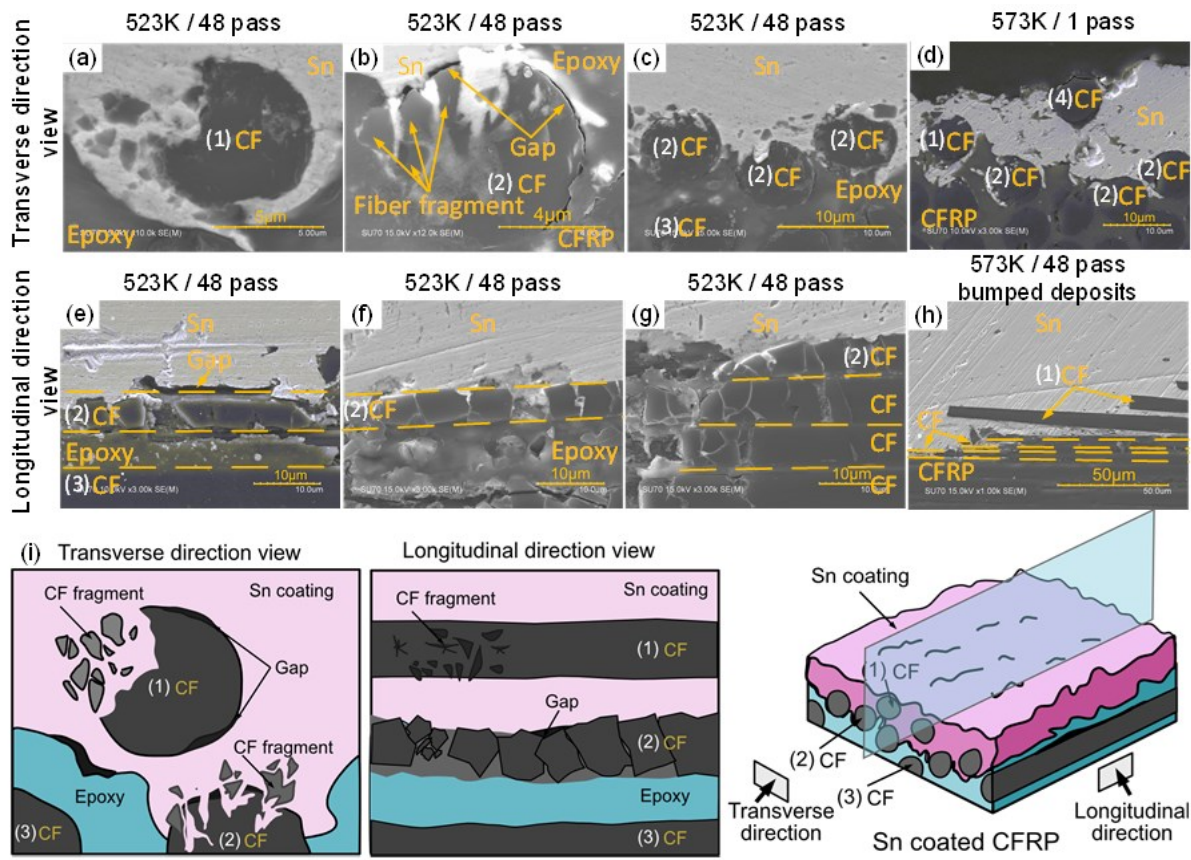

**Figure 5**. Sn/CF joint can Sn/CF/epoxy joint under 523K and 573K processing conditions (a) - (c) transverse direction view under 523 K/ 48 pass condition, (d) transverse direction view under 523 K/ 1 pass condition, (e) - (f) longitudinal direction of CF breakage performance under 523 K/ 48 pass condition, (h) longitudinal

direction view of bumped deposits under 523 K/ 48 pass condition, and (i) schematic figure of CF performance.

Sn/CF joint and Sn/CF/epoxy joint occasionally occurred at 523 K, but they are main combinations under 573 K case instead. Four carbon fragment performances can be investigated according to their position regarding the interface of joint type, which can be illustrated in Fig. 5(a) - (h). Combining the schematic figure in Fig. 5(i), "(1) CF" means whole CF merged inside the Sn coating, their interface is Sn/CF joint; "(2) CF" represents the upper part of CF embedded in the coating, but bottom part also remains in epoxy; "(3) CF" refers to the whole CF hinged in epoxy, the original state after epoxy cured during the CFRP manufacturing process. In Fig. 5(a), "(1) CF" shows that Sn coating wraps completely whole embedded CFs regardless of them experienced fracture or breakage; its fragments were randomly distributed adjacent to the main part of the broken fiber. In the condition of "(2) CF", shown in Fig. 5(b) and (c), along the transverse direction, the mixture of Sn and epoxy are inserted into cracks of carbon by subsequent particles, CF contact with epoxy and coating simultaneously. Usually, at the interface of Sn and single fiber, gaps are observed. Moreover, the broken carbon fibers are distributed along the longitudinal direction, the upper surface of the upper CF are bonded with the coating; also coating is investigated filled in the cracks of neighboring pieces, which promotes bonding capability. Fig. 5(e) - (9) shows the longitudinal direction view of "(2) CF", Sn/CF joint, and Sn/CF/epoxy joint are more common combinations, fractured epoxy and broken CFs compacted at the interface of uppermost substrate and coating. Probably because they were compressed by the Sn coating at the moment of breaking, and their irregular fragments caused the gap (Fig. 5(b, e)). Moreover, we observed the longitudinal view of the bumped deposits under 573K; severe surface corrosion causes CFs to be inserted inside of the coating. Besides, "(4) CF" represents the CFs exposure in the air passing through the whole coating, usually occur because sparse and thin coating at 573 K cannot completely cover its eroded substrate surface. For further exploring the temperature effects on the interface and assisting us in explaining the bonding mechanism, we analyzed the distribution of the surface temperature.

*3.4 Surface temperature distribution*

In order to fully understand and explain the thermal effects of every gas temperature, surface temperature distributions were conducted.

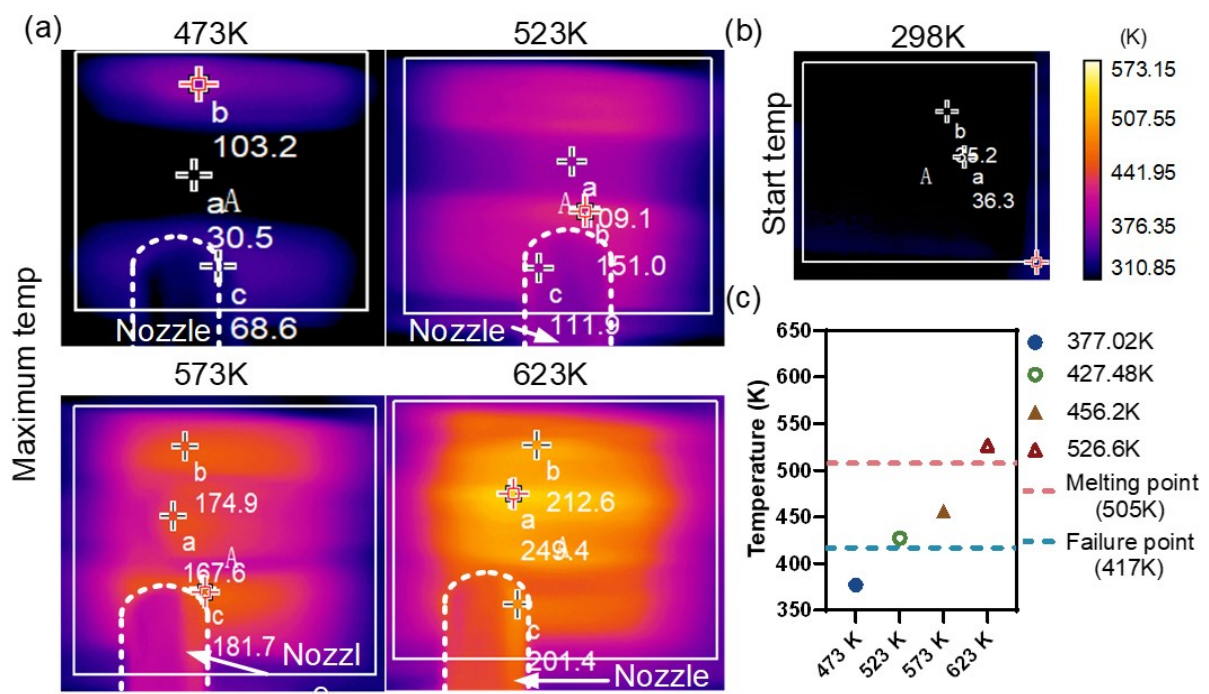

**Figure 6**. Surface temperature distributions during the spray process (a) temperature distributions where the maximum temperature was obtained, (b) start temperature distributions, and (c) maximum temperature.

Figure 6(b) shows the start temperature of RT before spraying; after spraying, the maximum temperature maps during the whole process under four gas temperatures are shown in Fig. 6(a). It should be noted that the unit of measured temperature marked on the maps is °C; the converted temperature can refer to the color bar in K.

In Fig. 6(b), with increasing the gas temperature from 473 K to 623 K, the maximum temperatures increase. The melting temperature of Sn of 505 K and failure temperature of epoxy of 417 K are marked by a red dot line and a blue dot line, respectively, which indicates only the 473 K case did not occur epoxy failure, only the 623 K occurred Sn melting phenomenon [34]. The analyses can provide evidence to verify surface morphology and afterward characterization of coating bottom surface and substrate surface after pulling off the coating.

*3.6. Adhesion strength measurement*

The pull-off adhesion strength tests of 48 passes multi-layer were conducted to verify the adhesion strength, and the distribution of Sn coating on the CFRP surface can further explicit the bonding mechanism. Under different processing conditions, the morphology of the substrate surfaces after pulling are various; two surface topographies of the bottom surface of the Sn coating and the upper surface of the substrate after puling are investigated in Fig. 7 and Fig. 8. In order to meet the requirement of the pull-off test, global coverage samples are prepared, as shown in Fig. 7(a).

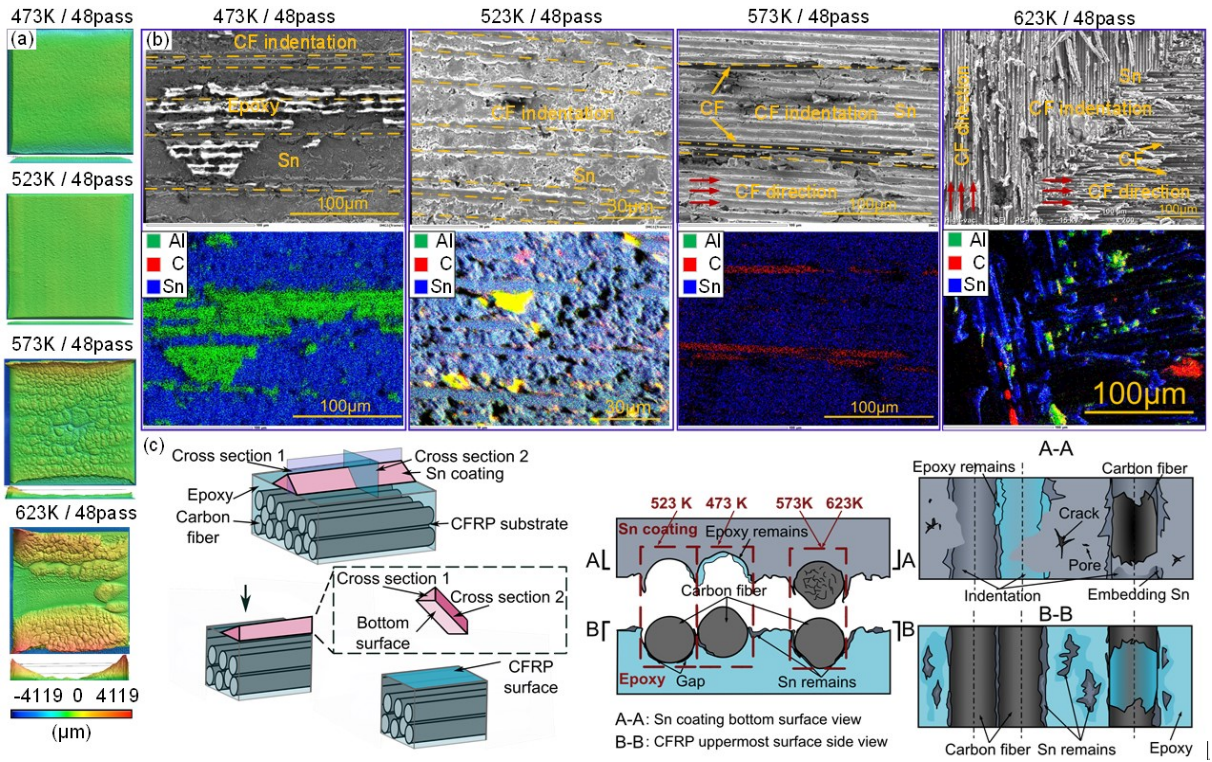

**Figure 7**. Coating bottom surface characteristics after the pull-off test under four gas temperatures, (a) global coverage samples, (b) coating bottom surface characteristics, (c) schematic figure of the pull-off test.

On the one hand, considering the coating surface, the CF indentation and convex performance can be illustrated in the SEM figure and corresponding schematic image in Fig. 7(b) and (c). On the condition of the gas temperature change from 473 K to 623 K, Sn particles plastically deformed, compressed, deposited, only melt at 623 K. Under 473 K case, epoxy resin remains and embeds in the bottom surface of Sn coating because 473 K cannot cause epoxy failure, which is consistent with surface temperature shown in Fig. 6(c). Hence, the indentations of carbon fibers marded by dotted lines can be observed both in the coating area and epoxy area. Almost no uppermost epoxy can be seen at 523 K and 573 K because epoxy began to degrade and decompose, driven by thermal effects. Meanwhile, more CF indentations shown may be indicated by two possible reasons. The one is , Sn/CF direct contact results from epoxy degradation; another is temperature increment. At 573 K and 623 K, unusual performance occurs: broken CFs attached and embedded in the Sn coating bottom surface. First, higher impact energy derived from higher temperature leads to CF fracture; afterward, the tighter combinations of coating and CFs or inserted CFs obtain higher adhesion strength, causing stronger sticking of CFs with Sn coating when pulling. When the temperature reaches 623 K, two continuous eroded laminates can be observed with very impacted and dense indentation, the warped part derives from CF-nether coating after the pull-off test. CFs drew out from surrounded coating, leaves one part connected, one part suspended. The schematic of this mechanism is illustrated in Fig. 7 (c).

It should be noted that the strip indentations are derived from upper compression of CFs, and flake convexes result from embedding coating in the vacancy of two adjacent CFs, and both sides of a single CF. Moreover, for any single fiber, the indentations and convexes are discontinuously located, the discontinuous areas can be regarded as the not tamping parts. Moreover, besides the fiber indentation area, another flat area was analyzed direct bonded with epoxy. The indentation has a smooth inner surface, which verifies carbon fiber cannot bond with Sn coating tightly, which indicates monarchical interlocking is weak. However, the flat area is rough and uneven, which means epoxy and Sn has a stronger binding force. Overall, a potential bonding mechanism can be investigated that with the increasing gas temperature, epoxy disappears, Sn gets more chance to deposit directly on the CF surface. Also, a higher temperature provides stronger adhesion strength seen in Fig. 7(d).

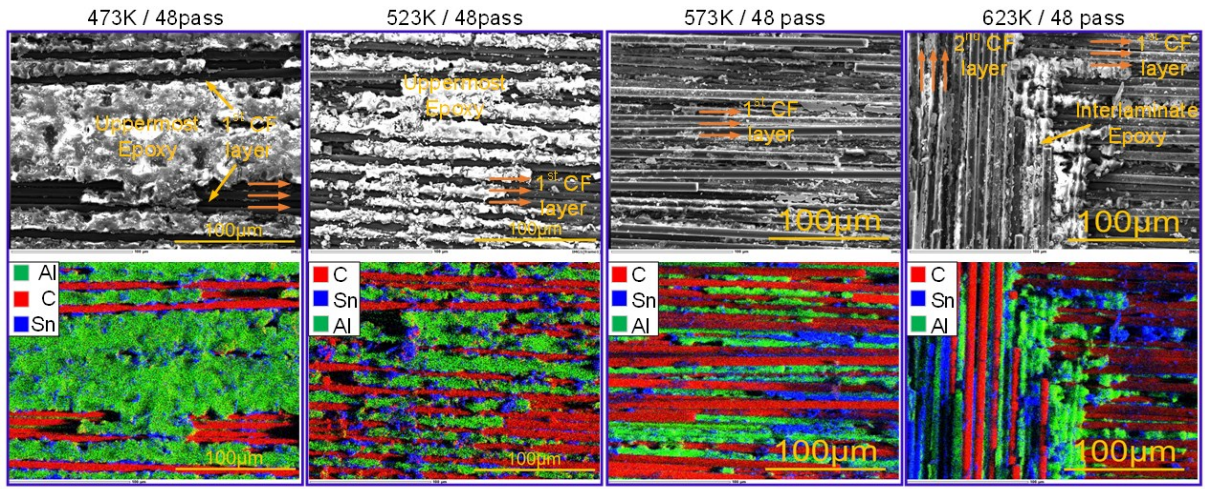

**Figure 8.** Substrate surface performance after the pull-off test

On the other hand, the substrate surfaces under various gas conditions after the pull-off test are observed shown in Fig. 8. When the gas temperature is 473 K, Sn coating residual epoxy and exposure CFs mainly distribute; exposure CFs showed because their epoxy cover layer is attached on the coating bottom surface corresponding with Fig. 7(b). Apparently, Sn coatings mainly distribute on the epoxy clusters or locate in vacancies between evenly arranged CFs. When increasing the temperature from 473K to 523K, epoxy gradually disappears, accompanying by more CFs exposure. Sn coating is distributed around the fiber, mainly scattered on the epoxy surface and embedded in the interface of fiber and epoxy; rare Sn coating locates on the fiber distribution area. Notably, comparing the probability occurrence of Sn/epoxy and Sn/CF joint, Sn/epoxy joint occupies more circumstances, which proves Sn particles are easier to bond with epoxy than CF. At 573 K, fully CF exposure and occasional breakage can be observed; while at 623 K, multi-laminates erosion occurs, at least two perpendicular CF directions of two laminates proves the serve erosion condition.

## 4. Discussion

## 4.1. Adhesion behavior analyses

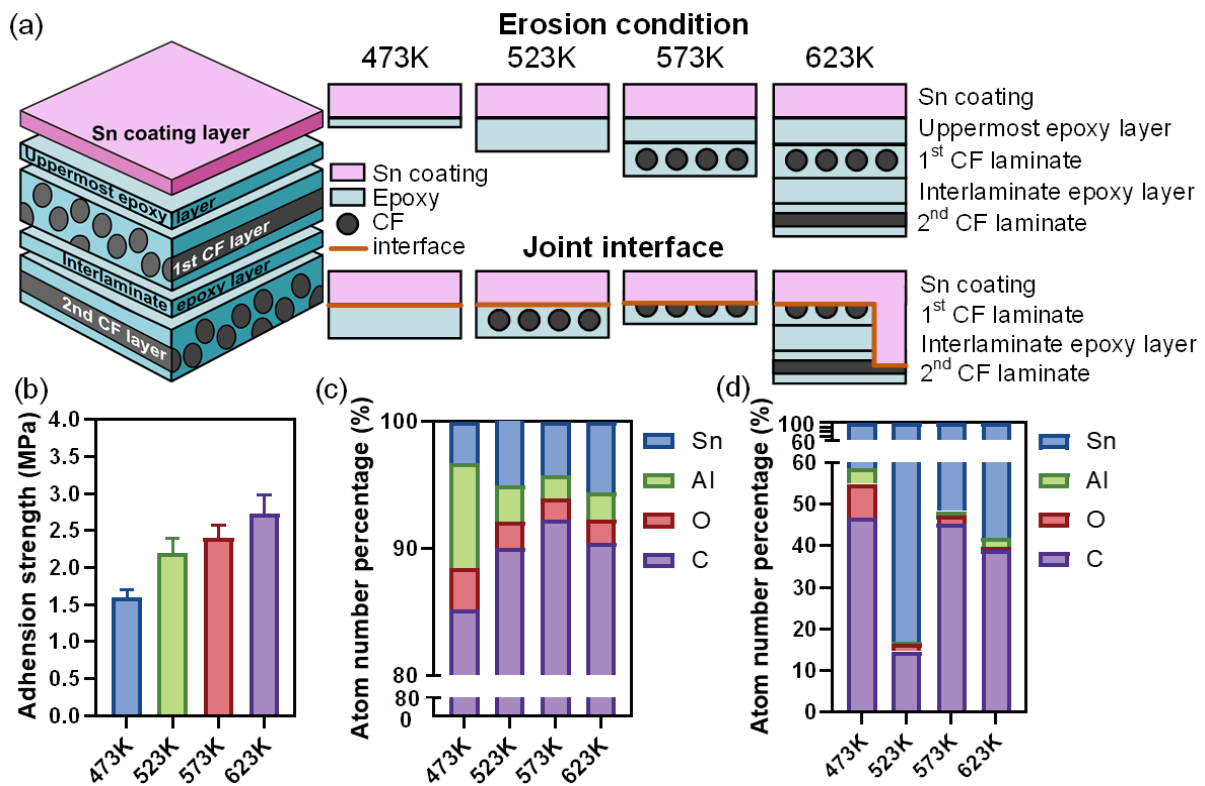

**Figure 9**. Erosion condition (a) CFRP structure, erosion condition and joint interface under various conditions, (b) adhesion strength, (c) atom number percentage of the substrate upper surface from EDX observation (d) atom number percentage of the coating bottom surface from EDX observation

From the microstructure observation of the two-side surface after the pull-off test, some interesting conclusions can be drawn. First, the structure of the coating bottom surface and the substrate upper surface correspond to each other. In Fig. 9(a), we specifically explain the structure of the CFRP sample again. Summarilly, under 473 K, almost no erosion of the first epoxy layer occurs, the interface obsdeved is Sn/epoxy joint; under 523 K, erosion happens in the first epoxy layer with Sn/CF laminate interface; under 573 K, erosion also occurs in the first epoxy layer involving the first CF laminate, while the interface is mainly Sn/CF interface, different from the situation of 523 K, direct contact of Sn/CF occupies the main part of combination; under 623K, above the second CF laminate are all the erosion regions, the interface consists of Sn/CF (longitudinal direction), Sn/epoxy (second layer) and Sn/CF (transverse direction). Second, the adhesion strength is shown in Fig. 9 (b); with the increasing temperature, the adhesion strength increases. These characteristics summarised under various temperatures illustrate when we only considering physical bonding, which is mechanical interlocking, but the bonding strength under 623K is the highest owing to metal melting. Actually, Sn particles can penetrate and embed into epoxy more easily than CF, fractured epoxy generates a rough surface, and various cracks and gaps help regardless of melted or solid Sn particles fill in. Meanwhile, more contact areas result from CF fragments, gaps between CF and epoxy, and even cracks inside the CF can further increase combined opportunity, enhance bonding effectiveness further promote deposition performance because the adhesion strength in the interface of Sn coating and CFRP substrate is weak [8,35]. Interestingly, Sn particles deform upon impact on the surface due to its softness nature, results in the jet formation and ensuing bond formation with the substrate via mechanical anchoring. Plastic deformation may disrupt thin surface films, provide intimate conformal contact under high local pressure permitting bonding to occur. Figure 9(c) and (d) are the figures of the ingredient percentage of atom number distribution measured by the EDX method after the pull-off test. Figure 9(c) shows the distribution of CFRP surface; the Carbon element (C) is mainly distributed in CF; Oxygen (O) and Aluminum element (Al) exist in epoxy, while Sn mainly represents the remained Sn coatings. Under 473 K, epoxy is not degraded so much, CFs is buried in epoxy resin. That's the reason for the minimum value of C and the maximum value of the O and Al element. The adhesion strength of Sn coating is worst in this condition, so the value of Sn is the minimum one among four circumstances. Increasing the temperature from 473 K to 523 K and 573 K, the distribution of carbon increases due to CF exposure, the representation of epoxy by O and Al decreases due to erosion and epoxy degradation. Sn increases under 523 K due to epoxy erosion but decreases due to CF exposure because Sn can better deposit on epoxy then CF surface. At 623 K, C decrease and Sn increase because the second CF laminate had been eroded accompanying with exposure of the second epoxy layer, resulting in an epoxy increment. Similarly, Figure 9(d) shows the element distribution of the coating bottom surface. Sn element occupies the main part, C element ranked second. Under 473 K, we obtain the most C element because we have the most epoxy (O and Al element). However, different from 473K, under 573K, C element is ranked second because of CF exposure; meanwhile, the value of epoxy is the minimum one. Under 523K, epoxy is almost gone with the lowest value of O and Al element, as well as the lowest C value. Under 623K, the second epoxy layer shows then increases the value of O and Al element.

### 4.3. Interface analyses

It is well known that the properties of the interface between two materials are governed by the chemical/morphological and physical/thermodynamic compatibility. The exposure of CF during the process improves the bonding condition.[9] The schematic figure of cold spray processing under high temperature (over 523K) is represented in Fig. 10 (a). As observed before, in the moment of high-temperature gas-particle Sn particles input, epoxy degraded, CFs exposed and suspended, particles formed wrapping with CFs, some particles rebonded beneath the CFs, some particles inserted between CFs, some particles bring in pressure on the CFs, then the embedded coating performance can be obtained.

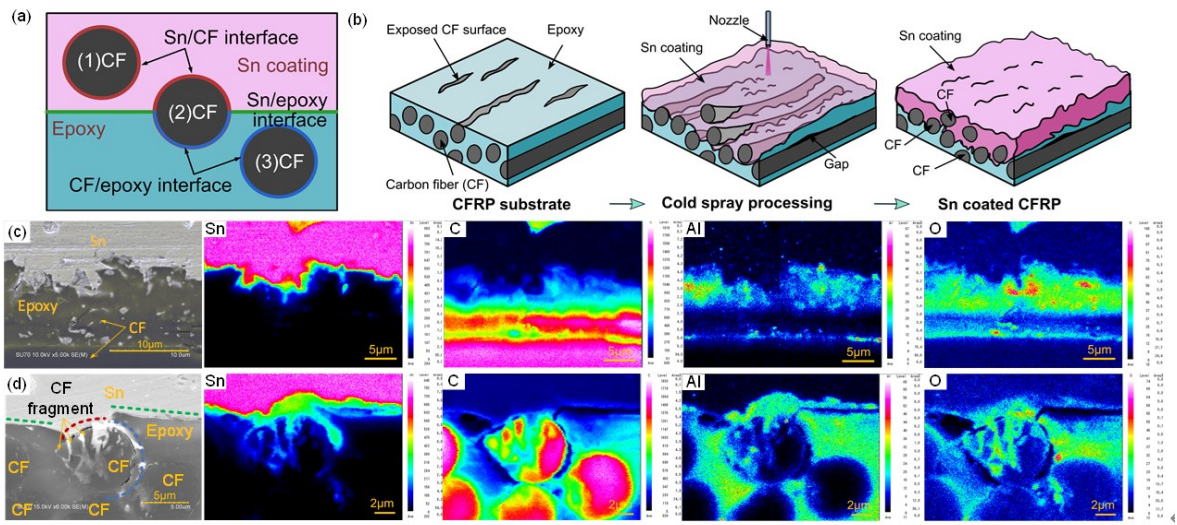

**Figure 10**. Interfacial element distribution and bonding mechanism explication (a) three kinds of CF-related interfaces, (b) schematic of deposition under the temperature higher than 523 K Interfacial (c) Sn/epoxy interlocked interface under 523 K condition (d) Sn/CF/epoxy interface under 523K condition

As mentioned aforehand in Fig. 9(a), interfacial morphologies are various corresponding to different processing conditions. Figure 10(a) represents three kinds of CF-related interface: Sn/epoxy interface is marked by a green dot line, Sn/CF interface is denoted by the red color, and CF/epoxy interfaces are presented by blue color. CFs circled by them are (1)CF, (2)CF, and (3)CF, respectively. Figure 10(b) illustrates the bonding mechanism under 573 K and 623 K occurring (1)CF feature. As shown, the original CFRP samples contain some degree of CF exposure due to low humidity resistance of epoxy. After spraying, epoxy erodes, CF exposes, the upper CF partially warp and partially break, melting Sn flows beneath the warped CF then consolidates; while, unmelted particles are impacted, deformed then fill in the gap and cracks, finally coating forms.

Figure 10(c) shows the Sn/epoxy interlocked interface obtained under the gas temperature of 523 K. There happens Sn atoms diffusion in the interfacial region towards epoxy. Further, Fig. 10(d) shows a complicated Sn/CF/epoxy joint circumstance contains three kinds of interfaces simultaneously. In this case, Sn coating inserted in the broken CF and contact with epoxy and CF to form Sn/epoxy and Sn/CF joint simultaneously. In Fig. 10(d), obvious gaps can be recognized in the Sn/CF interface (marked in red) rather than Sn/epoxy (marked in green) interface from the EMPA observation, indicates Sn/epoxy is tighter and stronger. It again indicates that Sn has better adhesion conditions with epoxy rather than CF.

## 5. Conclusions

Cold sprayed Sn coating onto CFRP substrate under four gas temperatures were carried out. The bonding mechanisms are elucidated under different conditions. The following conclusions were made based on the results.

1. From the morphology of surface and interface, there exist six patterns of interface performances under four conditions. In general, it can be classified into three categories: Sn/epoxy joint, Sn/CF joint, and Sn/CF/epoxy joint. At 473 K, only Sn/epoxy occurs. Sn/CF/epoxy combination occurs under the other three temperature conditions, but mainly under 523 K. Instead, the Sn/CF joint only occurs at 573 K and 623 K because of epoxy erosion. Epoxy is mainly shown in lower temperatures (473K); CFs are mainly observed in higher temperatures (623K). Meanwhile, epoxy and CF are both presented in middle temperatures (523K and 573K).

2. When increasing the gas temperature from 473K to 623K, epoxy gradually degrads, CF laminates gradually fractures, CF laminates gradually exposes then wap and break. Laminate erosion happens at 573 K, multi-laminate erosion becomes severe at 623 K with two continuous CF laminates are eroded in our case.
3. Sn/epoxy has better bonding behavior than Sn/CF joint due to existing of gaps in the Sn/CF interfaces affecting the bonding quality. This conclusion provides evidence and new ideas that are, increasing Sn/epoxy bonding ratio to ensure epoxy gets more chance to contact Sn coating before erosion.

**Acknowledgment**

The authors thank Hajime Yoshinaga and Tsuyoshi Oguri, JAMCO Corporation, for providing CFRP samples. This research was supported by the NEDO Feasibility Study Program "Multi-material 3D joining and optimum molding technology for higher reliability than the current joining in the aviation field".